\documentclass[twocolumn,twocolappendix]{aastex631}
\usepackage{graphicx}

\begin{document}

\title{The day-long, repeating GRB\,250702BDE / EP250702a: A unique extragalactic transient}
\shorttitle{The extragalactic repeating GRB\,250702BDE}
\shortauthors{Levan et al.}

\correspondingauthor{Andrew J. Levan, Antonio Martin-Carrillo}
\email{a.levan@astro.ru.nl, antonio.martin-carrillo@ucd.ie}

\author[0000-0001-7821-9369]{Andrew J. Levan}
\affiliation{Department of Astrophysics/IMAPP, Radboud University, PO Box 9010, 6500 GL Nijmegen, The Netherlands}

\author[0000-0001-5108-0627]{Antonio Martin-Carrillo}
\affiliation{School of Physics and Centre for Space Research, University College Dublin, Belfield, Dublin 4, Ireland}

\author[0000-0003-1792-2338]{Tanmoy Laskar}
\affiliation{Department of Physics \& Astronomy, University of Utah, Salt Lake City, UT 84112, USA}

\author[0000-0002-8775-2365]{Rob A. J. Eyles-Ferris}
\affiliation{School of Physics and Astronomy, University of Leicester, Leicester, LE1 7RH, UK}

\author[0000-0002-5460-6126]{Albert Sneppen}
\affiliation{Niels Bohr Institute, University of Copenhagen, Jagtvej 155, 2200, Copenhagen N, Denmark}
\affiliation{The Cosmic Dawn Centre (DAWN), Denmark}

\author[0000-0003-3193-4714]{Maria Edvige Ravasio}
\affiliation{Department of Astrophysics/IMAPP, Radboud University, PO Box 9010, 6500 GL Nijmegen, The Netherlands}
\affiliation{INAF-Osservatorio Astronomico di Brera, Via Bianchi 46, I-23807, Merate (LC), Italy}

\author[0000-0002-9267-6213]{Jillian C.~Rastinejad}
\affiliation{Center for Interdisciplinary Exploration and Research in Astrophysics (CIERA) and Department of Physics and Astronomy, Northwestern University, Evanston, IL 60208, USA}

\author[0000-0002-7735-5796]{Joe S. Bright}
\affiliation{Astrophysics, Department of Physics, University of Oxford, Keble Road, Oxford OX1 3RH, UK}

\author[0000-0002-0426-3276]{Francesco Carotenuto}
\affiliation{INAF-Osservatorio Astronomico di Roma, Via Frascati 33, I-00078, Monte Porzio Catone (RM), Italy}

\author[0000-0001-9842-6808]{Ashley A. Chrimes}
\affiliation{European Space Agency (ESA), European Space Research and Technology Centre (ESTEC), Keplerlaan 1, 2201 AZ Noordwijk, the Netherlands}
\affiliation{Department of Astrophysics/IMAPP, Radboud University, PO Box 9010, 6500 GL Nijmegen, The Netherlands}

\author[0009-0009-1573-8300]{Gregory Corcoran}
\affiliation{School of Physics and Centre for Space Research, University College Dublin, Belfield, Dublin 4, Ireland}

\author[0000-0002-5826-0548]{Benjamin P. Gompertz}
\affiliation{School of Physics and Astronomy and Institute for Gravitational Wave Astronomy, University of Birmingham, Birmingham B15 2TT, UK}

\author[0000-0001-5679-0695]{Peter G. Jonker}
\affiliation{Department of Astrophysics/IMAPP, Radboud University, PO Box 9010, 6500 GL Nijmegen, The Netherlands}

\author[0000-0001-5169-4143]{Gavin P. Lamb}
\affiliation{Astrophysics Research Institute, Liverpool John Moores University, IC2 Liverpool Science Park, 146 Brownlow Hill, Liverpool, L3 5RF, UK}

\author[0000-0002-7517-326X]{Daniele B. Malesani}
\affiliation{Niels Bohr Institute, University of Copenhagen, Jagtvej 155, 2200, Copenhagen N, Denmark}
\affiliation{The Cosmic Dawn Centre (DAWN), Denmark}

\author[0000-0002-6950-4587]{Andrea Saccardi}
\affiliation{Universit\'e Paris-Saclay, Universit\'e Paris Cit\'e, CEA, CNRS, AIM, 91191, Gif-sur-Yvette, France}

\author[0000-0003-2276-4231]{Javier Sánchez Sierras}
\affiliation{Department of Astrophysics/IMAPP, Radboud University, PO Box 9010, 6500 GL Nijmegen, The Netherlands}

\author[0000-0003-4876-7756]{Benjamin Schneider}
\affiliation{Aix Marseille Univ., CNRS, CNES, LAM, Marseille, France}

\author[0000-0001-6797-1889]{Steve Schulze}
\affiliation{Center for Interdisciplinary Exploration and Research in Astrophysics (CIERA), Northwestern University, 1800 Sherman Ave., Evanston, IL 60201, USA}

\author[0000-0003-3274-6336]{Nial R. Tanvir}
\affiliation{School of Physics and Astronomy, University of Leicester, Leicester, LE1 7RH, UK}

\author[0000-0001-9398-4907]{Susanna D. Vergani}
\affiliation{LUX, Observatoire de Paris, Universit\'e PSL, CNRS, Sorbonne Universit\'e, Meudon, 92190, France}

\author[0000-0002-4465-8264]{Darach Watson}
\affiliation{Niels Bohr Institute, University of Copenhagen, Jagtvej 155, 2200, Copenhagen N, Denmark}
\affiliation{The Cosmic Dawn Centre (DAWN), Denmark}

\author[0009-0000-5068-3434]{Jie An}
\affiliation{National Astronomical Observatories, Chinese Academy of Sciences, Beijing 100012, China}
\affiliation{School of Astronomy and Space Science, University of Chinese Academy of Sciences, Chinese Academy of Sciences, Beijing 100049, China}

\author[0000-0002-8686-8737]{Franz E. Bauer}
\affiliation{Instituto de Alta Investigaci{\'{o}}n, Universidad de Tarapac{\'{a}}, Casilla 7D, Arica, Chile}

\author[0000-0001-6278-1576]{Sergio Campana}
\affiliation{INAF-Osservatorio Astronomico di Brera, Via Bianchi 46, I-23807, Merate (LC), Italy}

\author[0000-0002-7910-6646]{Laura Cotter}
\affiliation{School of Physics and Centre for Space Research, University College Dublin, Belfield, Dublin 4, Ireland}

\author[0009-0007-6927-7496]{Joyce N.D. van Dalen}
\affiliation{Department of Astrophysics/IMAPP, Radboud University, PO Box 9010, 6500 GL Nijmegen, The Netherlands}

\author[0000-0003-3703-4418]{Valerio D'Elia}
\affiliation{Space Science Data Center (SSDC) - Agenzia Spaziale Italiana (ASI), Via del Politecnico snc, I-00133 Roma, Italy}

\author[0000-0002-4036-7419]{Massimiliano De Pasquale}
\affiliation{MIFT Department, University of Messina, Via F. S. D'Alcontres 31, 98166, Messina, Italy.}

\author[0000-0001-7717-5085]{Antonio de Ugarte Postigo}
\affiliation{Aix Marseille Univ., CNRS, CNES, LAM, Marseille, France}

\author[0000-0001-9868-9042]{Dimple}
\affiliation{School of Physics and Astronomy and Institute for Gravitational Wave Astronomy, University of Birmingham, Birmingham B15 2TT, UK}

\author[0000-0002-8028-0991]{Dieter H. Hartmann}
\affiliation{Clemson University, Department of Physics and Astronomy, Clemson, SC 29634, USA}

\author[0000-0002-4571-2306]{Jens Hjorth}
\affiliation{DARK, Niels Bohr Institute, University of Copenhagen, Jagtvej 155A, 2200 Copenhagen, Denmark}

\author[0000-0001-9695-8472]{Luca Izzo}
\affiliation{INAF, Osservatorio Astronomico di Capodimonte, Salita Moiariello 16, I-80121 Naples, Italy}
\affiliation{DARK, Niels Bohr Institute, University of Copenhagen, Jagtvej 155A, 2200 Copenhagen, Denmark}

\author[0000-0002-9404-5650]{P\'all Jakobsson}
\affiliation{Centre for Astrophysics and Cosmology, Science Institute, University of Iceland, Dunhagi 5, 107 Reykjavik, Iceland}

\author[0000-0002-4870-9436]{Amit Kumar}
\affiliation{Department of Physics, Royal Holloway - University of London, Egham, TW20 0EX, UK} 

\author[0000-0002-2810-2143]{Andrea Melandri}
\affiliation{INAF-Osservatorio Astronomico di Roma, Via Frascati 33, I-00078, Monte Porzio Catone (RM), Italy}

\author[0000-0002-5128-1899]{Paul O'Brien}
\affiliation{School of Physics and Astronomy, University of Leicester, Leicester, LE1 7RH, UK}

\author[0000-0002-8875-5453]{Silvia Piranomonte}
\affiliation{INAF-Osservatorio Astronomico di Roma, Via Frascati 33, I-00078, Monte Porzio Catone (RM), Italy}

\author[0000-0003-3457-9375]{Giovanna Pugliese}
\affiliation{Anton Pannekoek Institute of Astronomy, University of Amsterdam, P.O. Box 94249, 1090 GE Amsterdam, The Netherlands}

\author[0000-0001-8602-4641]{Jonathan Quirola-V\'asquez}
\affiliation{Department of Astrophysics/IMAPP, Radboud University, PO Box 9010, 6500 GL Nijmegen, The Netherlands}

\author[0000-0001-5803-2038]{Rhaana Starling}
\affiliation{School of Physics and Astronomy, University of Leicester, Leicester, LE1 7RH, UK}

\author[0000-0003-0121-0723]{Gianpiero Tagliaferri}
\affiliation{INAF-Osservatorio Astronomico di Brera, Via Bianchi 46, I-23807, Merate (LC), Italy}

\author[0000-0003-3257-9435]{Dong Xu}
\affiliation{National Astronomical Observatories, Chinese Academy of Sciences, Beijing 100012, China}

\author[0009-0009-8473-3407]{Makenzie E. Wortley}
\affiliation{School of Physics and Astronomy and Institute for Gravitational Wave Astronomy, University of Birmingham, Birmingham B15 2TT, UK}


\begin{abstract}
$\gamma$-ray bursts (GRBs) are singular outbursts of high-energy radiation with durations typically lasting from milliseconds to minutes and, in extreme cases, a few hours. They are attributed to the catastrophic outcomes of stellar-scale events and, as such, are not expected to recur. Here, we present observations of an exceptional GRB\,250702BDE which triggered the {\em Fermi} gamma-ray burst monitor on three occasions over several hours, and which was detected in soft X-rays by the \textit{Einstein Probe} a day before the $\gamma$-ray triggers (EP250702a). We present the discovery of an extremely red infrared counterpart of the event with the VLT, as well as radio observations from MeerKAT. Hubble Space Telescope observations pinpoint the source to a non-nuclear location in a host galaxy with complex morphology, implying GRB 250702BDE is an extragalactic event. The multi-wavelength counterpart is well described with standard afterglow models at a relatively low redshift $z \sim 0.2$, but the prompt emission does not readily fit within the expectations for either collapsar or merger-driven GRBs. Indeed, a striking feature of the multiple prompt outbursts is that the third occurs at an integer multiple of the interval between the first two. Although not conclusive, this could be indicative of periodicity in the progenitor system. We discuss several possible scenarios to explain the exceptional properties of the burst, which suggest that either a very unusual collapsar or the tidal disruption of a white dwarf by an intermediate-mass black hole are plausible explanations for this unprecedented GRB. 
\end{abstract}


\section{Introduction}

$\gamma$-ray bursts (GRBs) are brief flashes of high-energy radiation (peaking at $\sim 1-1000$ keV) with typical durations spanning from a fraction of a second (short GRBs), to minutes (long GRBs) \citep{kouveliotou93}, with only a tiny minority having durations up to a few hours (so-called ultra-long GRBs; \citealt{levan14}). A key feature of GRBs is that they are singular, non-repeating events that represent the final moments of stars, either via the collapse of a stellar core \citep[e.g.][]{hjorth03,stanek03} or the merger of two compact objects \citep[e.g.][]{tanvir13,berger13,abbott_bns}. 

Although there is a great deal of diversity in individual GRB light curve shapes, bursts at the extremes in duration or variability are rare and represent novel opportunities to search for GRB progenitors outside the accepted paradigm. For example, a subpopulation of the very shortest bursts arises from repeating sources -- flares from magnetars, either within the Milky Way, or beyond \citep{hurley05,burns21,Mereghetti+24,Trigg+25}. In several instances, $\gamma$-ray emission has been indicative of relativistic outflows from a tidal disruption event (TDE; \citealt{burrows11,levan11,bloom11,Cenko2012a,Brown2015a}). However, in the case of TDEs, the related $\gamma$-ray light curves do not resemble GRBs, being longer lived, but much less ``bursty''\footnote{The $\gamma$-ray discovered events were located by searching for $\gamma$-ray triggers in reconstructed images rather than via the typical rate triggers.}. Strikingly, although only a handful of GRB progenitors have been identified observationally, there are a large number of plausible routes that could lead to $\gamma$-ray emission, including events of significant astrophysical importance such as accretion-induced collapse, \citep{nomoto91}; disruptions of a white dwarf (WD) by an intermediate mass black hole (IMBH, $\sim10^3-10^5 M_\odot$), \cite{irwin10}); micro-TDEs, \citep{perets2016}; other stellar mergers, \citep{fryer99,deMink14}; or even explosions within common envelopes \citep{fryer99,schroder20}. Cases of rare GRBs therefore provide a novel route to identify some of the rarest, but most astrophysically important events in nature. 

Here, we present observations of a recently discovered series of GRBs (GRBs\,250702D, 250702B and 250702E) which are unique in both their temporal and spatial coincidence and in their multi-wavelength properties. We explore several progenitor scenarios for this series of GRBs and assess their viability. Throughout this work, we report all magnitudes in the AB system and assume a cosmology of $\Lambda$CDM world model with $\Omega_M = 0.315$, $\Omega_{\Lambda} = 0.685$, and $H_0 = 67.4$ km s$^{-1}$ Mpc$^{-1}$ \citep{planck18}.

\begin{figure*}
\centering
\includegraphics[angle=0,width=\textwidth]{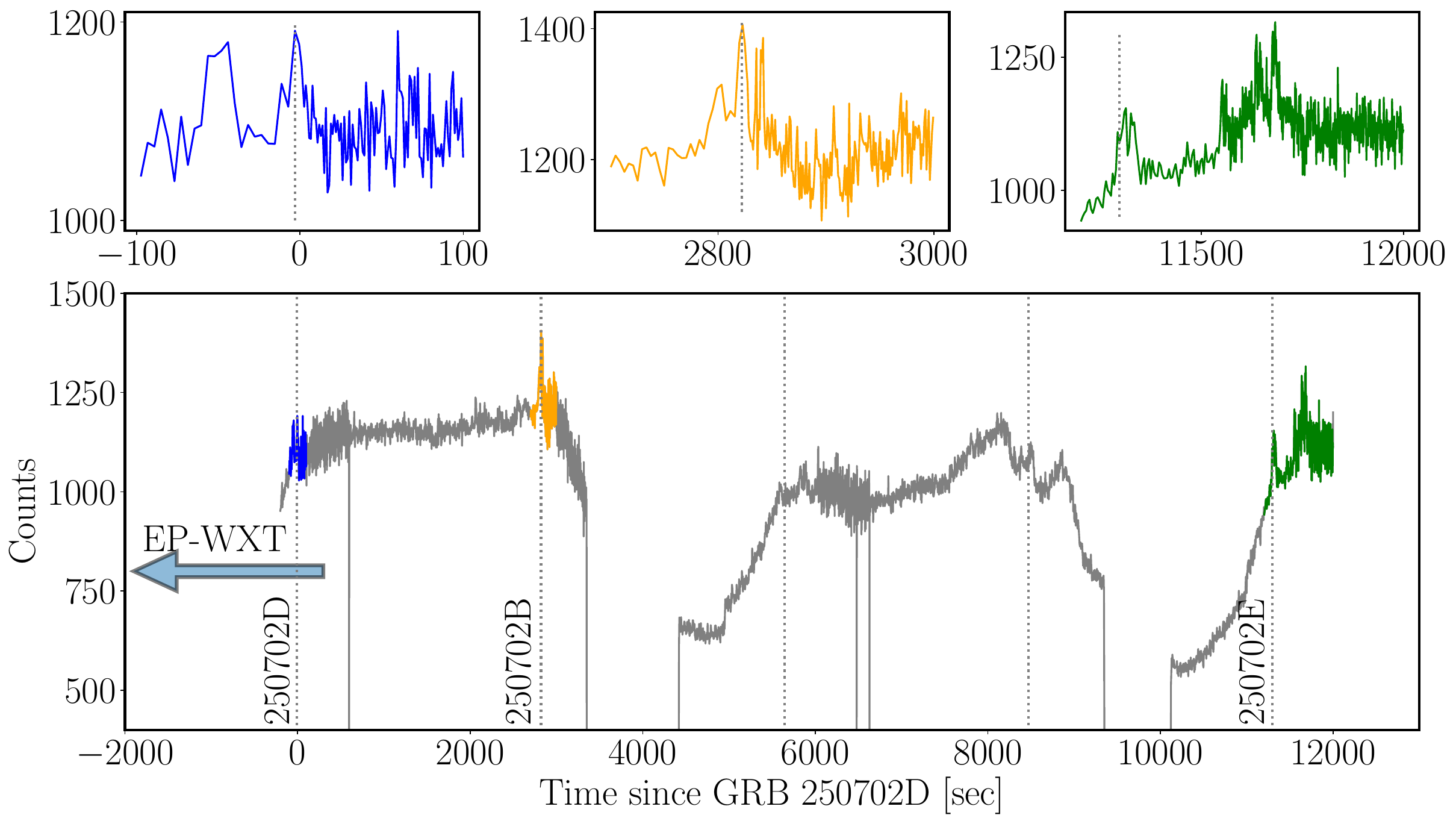}
\caption{The {\em Fermi}-GBM (8\,--\,900\,keV energy range; observer frame) count-rate light curve spanning from $-2\times10^3$ to $1.2\times10^4$~s after the GRB\,250702D trigger (main panel, grey lines; with data from each burst highlighted in color), along with zoom-ins around the time of the individual triggers (upper row). EP-WXT observations revealed an X-ray counterpart a day earlier (blue arrow; \citealt{gcn40906}), indicating transient high-energy emission preceding the GBM trigger. The dashed vertical lines indicate periods of 2825s (the measured gap between GRB 250702D and GRB 250702B).
}
\label{fig:fermi_prompt}
\end{figure*}

\section{Observations} 

\subsection{Gamma-ray Detections}

GRBs\,250702D, 250702B and 250702E (hereafter ``GRB\,250702BDE'') were first discovered as a series of discrete triggers by the {\em Fermi} Gamma-Ray Burst Monitor (GBM; \citealt{gcn40891}) on July 2 at 13:09:02, 13:56:06, and 16:21:33 UT, respectively\footnote{Note that the alphabetical trigger names do not follow chronological order.}. Throughout this work, $\delta t$ refers to the time since the trigger time ($T_0$) of GRB\,250702D. A fourth event (GRB\,250702C) identified during this period was determined to be unrelated to GRB\,250207BDE due to its position on the sky \citep{gcn40931}. Konus-Wind also detected coincident hard X-rays (20--1250\,keV, observer frame) emission spanning this period \citep{gcn40914}. 
Coincident prompt $\gamma$-ray to hard X-ray emission was also detected by \textit{Swift}-BAT GUANO \citep{Tohuvavohu+20,GCN250702_BAT}, MAXI/GSC \citep{gcn40910}, GECAM-B \citep{gcn40922}, and the Space Variable Objects Monitor (SVOM) Gamma-ray Monitor (GRM; \citealt{gcn40923}). We extract the {\em Fermi}/GBM $\gamma$-ray light curve of all three GRB triggers in the energy range 8-900 keV from the daily data\footnote{\url{https://heasarc.gsfc.nasa.gov/W3Browse/fermi/fermigdays.html}} using the GBM Data Tools~\citep{GbmDataTools} and present it in Figure~\ref{fig:fermi_prompt}. 

\subsection{X-ray Observations}
\label{text:xray}

An X-ray transient associated with GRB\,250702BDE was first reported by the \textit{Einstein Probe} (EP) Wide Field X-ray Telescope (WXT; EP\,250702a, \citealt{gcn40906}). Subsequent stacking of WXT observations revealed that EP\,250702a was first detected on 1 July 2025, a day prior to the GRB detections \citep{gcn40906}. 
Observations by {\em Swift}-XRT beginning at $\delta t = 0.517$~days detected a bright, highly variable counterpart and provided the best X-ray position of R.A. (J2000)\,=\,18h\,58m\,45.61s, Dec. (J2000)\,=\,-07d\,52\arcmin\,26.9$^{\prime \prime}$ \citep{gcn40919} with a $2\arcsec$ uncertainty\footnote{The \emph{Swift}/XRT data is automatically processed as described in \cite{evans07, evans09} and can be found at \url{https://www.swift.ac.uk/LSXPS/transients/9377}.}. 
Additional follow-up observations with \textit{NuSTAR} were obtained starting at $\delta t = 1.317$~days and clearly detect a variable X-ray counterpart at 3-79 keV \citep{gcn41014}.

\subsection{Optical and Near-IR Observations}
\label{text:data:opt}
\begin{figure*}
\centering
\includegraphics[angle=0,width=17cm]{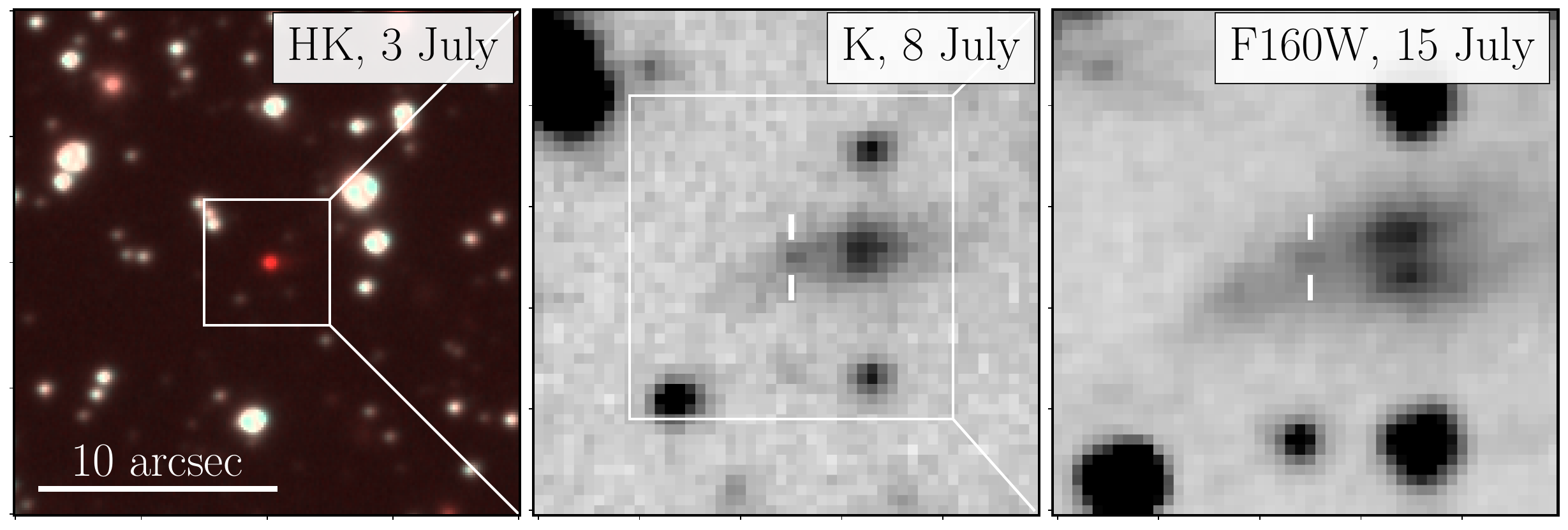}
\caption{Left: Discovery image of the counterpart of GRB\,250702BDE obtained with the VLT/HAWK-I. The false colour composite is constructed from the $H$- and $K$-band observations and demonstrates the transient's extremely red colour. Middle: Zoomed-in region (corresponding to the white box in the left panel) around GRB\,250702BDE's counterpart at $\delta t = 5.737$~days as observed with HAWK-I in excellent seeing conditions. The transient appears to be offset from a likely extended source. Right: \textit{HST} imaging clearly resolves the extended source, revealing a complex, asymmetric morphology, as well as the possible presence of a strong dust lane through the disk of the galaxy. }
\label{fig:imaging}
\end{figure*}

We initiated optical and near-IR observations of GRB\,250702BDE with HAWK-I on the Very Large Telescope (VLT, program ID 114.27PZ; PIs Malesani, Tanvir, Vergani) starting on 2025 July 3 at 07:03:03 UT ($\delta t = 0.746$~days). We process all VLT observations using the ESO \texttt{REFLEX} packages, astrometrically align the HAWK-I observations to the \textit{Gaia} reference frame, and perform photometric calibration with the UKIDSS Galactic Plane Survey \citep{Lucas+08}. We present a log of the HAWK-I observations in Table~\ref{tab:phot}.

In our first epoch of HAWK-I imaging, we clearly detect a new source in both $H$- and $K$-band within the {\em Swift}-XRT localization at R.A. (J2000)\,=\,18h\,58m\,45.57s, Dec. (J2000)\,=\,-07d\,52\arcmin\,26.2\arcsec, with an uncertainty of $\sim 0.1$\arcsec\, in each coordinate (Figure~\ref{fig:imaging}). We perform photometry on the source using small (0.5\arcsec) apertures, giving $K = 19.36 \pm 0.02$ mag, substantially brighter than the limiting magnitudes of archival VISTA observations of the field. We thus identify the source as the infrared counterpart of GRB\,250702BDE \citep{gcn40924}. Notably, in our first epoch of imaging, the source shows an extremely red colour of $H\,-\,K\,\rm{(AB)}\,=\,1.42\,\pm\,0.06$~mag (\citealt{gcn40961}; 1.37~mag when considering the Milky Way extinction of A$_{\rm V} = 0.847$~mag in the direction of the near-IR counterpart, \citealt{SF11}), corresponding to an extremely steep spectral index of $\beta\approx-4$ (with $F_\nu\propto t^\alpha\nu^{\beta}$). In addition to the transient point source, our observations reveal potential underlying extension in the E-W direction, consistent with an underlying host galaxy (Figure~\ref{fig:imaging}). In later epochs of HAWK-I imaging, we continue to use narrow (0.5\arcsec) apertures to minimize host contamination, which increases in later epochs, in the photometry. We therefore also provide magnitudes from subtraction of the final HAWK-I epoch at $\delta(t) \sim 12.5$ days, although we note that there is still likely transient light in the template epoch and the resulting magnitudes may be underestimated.

\begin{figure*}
\centering
\includegraphics[angle=0,width=14cm]{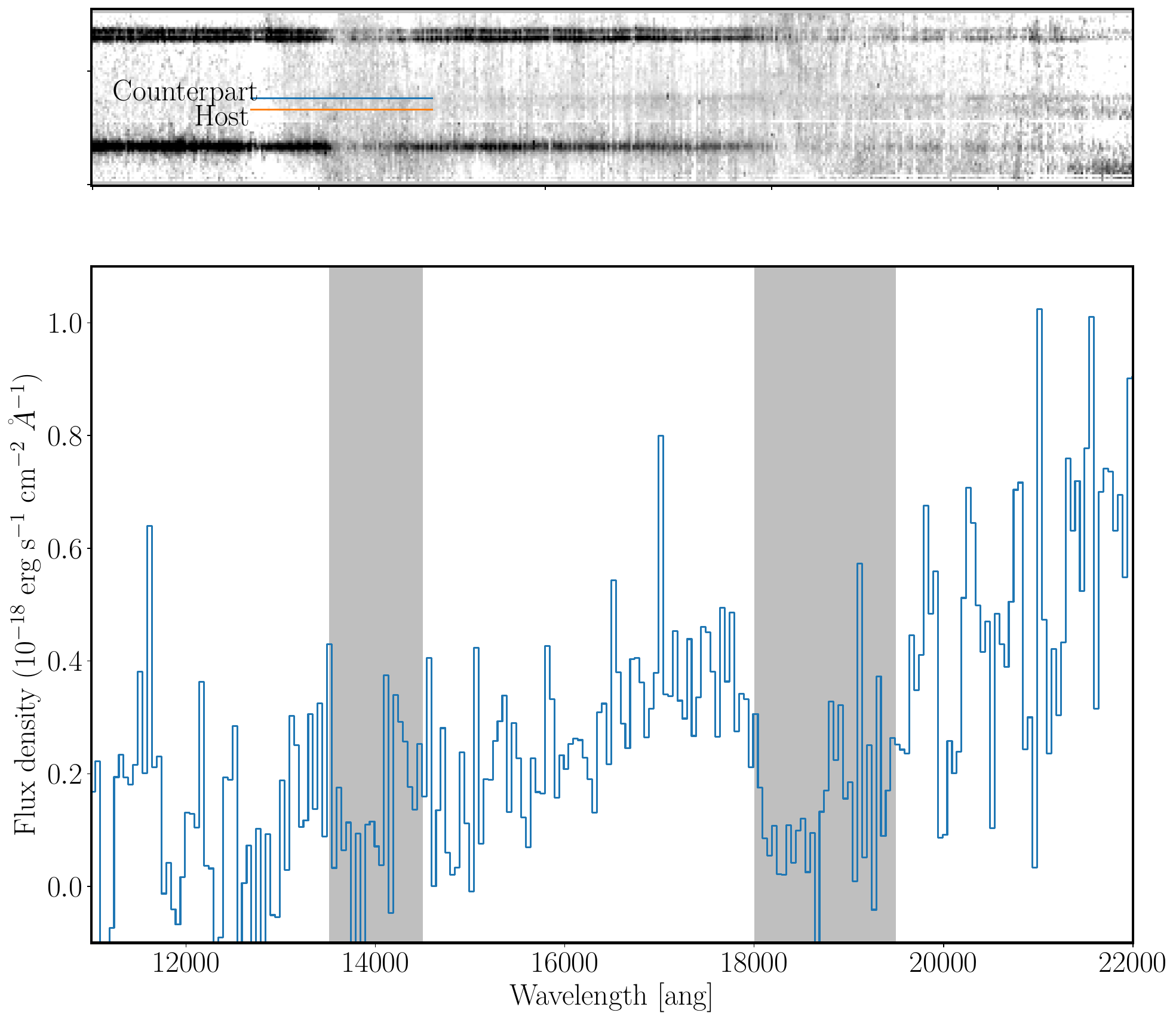}
\caption{X-shooter spectroscopy in 2D (top) and 1D (bottom) of the counterpart of GRB\,250702BDE obtained on 4 July 2025, rebinned in the 1D to $50$\,\AA\,per bin. Signal is only recovered in the infrared arm (shown) at low signal to noise. The grey bands indicate regions of high telluric absorption that are not corrected for. The source shows brightening through the H and K regions of the spectrum consistent with photometry. Given the low signal to noise no significant absorption or emission features are visible in the counterpart spectrum. The slit was also oriented to cover the location of the host galaxy, but no redshift could be derived.}
\label{fig:spectroscopy}
\end{figure*}

Following the HAWK-I near-IR detection, we obtained $4 \times 600$~s spectroscopy of the counterpart starting on 2025 July 4 at 03:26:53 UT ($\delta t = 1.60$~days) with the VLT X-shooter spectrograph \citep{Vernet2011}, covering 3\,000\,--\,24\,000\,\AA.  We oriented the slit with a position angle of 90 degrees to cover both the counterpart and the potential host galaxy, and employed the ABBA nod-on-slit mode. We reduce each individual exposure of the near-IR arm using the STARE mode reduction in the ESO reflex packages (a nodding reduction was not possible because of the presence of other objects in the slit). We subtract sky features and individually flux-calibrate each spectrum before combining into a final science spectrum \citep{Selsing19}. The resulting 2D spectra shows only a very faint trace across the $H$ and $K$-bands, highlighting the very red color of the transient (Figure~\ref{fig:spectroscopy}) and the low signal-to-noise ratio of the data does not enable any useful constraints on the redshift of either the counterpart or the putative host galaxy.

We furthermore initiated a request for Directors Discretionary Time (DDT) observations with the Hubble Space Telescope (\textit{HST}; program 17988, PI: Levan), and obtained a single orbit of observations with the F160W filter. We download the flat-fielded, dark-subtracted and CTE-corrected image from the MAST archive, and redrizzled the individual exposures to a final pixel scale of 0.07\arcsec. We detect both the near-IR counterpart as well as the underlying galaxy in the F160W image (Figure~\ref{fig:imaging}). The magnitude of the transient is highly uncertain because of contamination by galaxy light. Using a narrow aperture and on-galaxy background we obtain F160W=24.8 $\pm$ 0.3~mag, but note that the uncertainty is likely underestimated due to systematics. We measure an offset between the point source and the galaxy centroid\footnote{We note that the galaxy centroid is not well defined, since it is divided into two visible regions in the \textit{HST} imaging (Figure~\ref{fig:imaging}). However, the offset from either to the GRB position is comparable.} of $0.7\arcsec$.

We also obtained observations with the Gran Telescopio Canarias (GTC) using both the IR imager (EMIR) and the simultaneous multi-band optical imager HIPERCAM, as well as with the Nordic Optical Telescope (NOT), in multiple bands. We reduce observations using standard instrument pipelines. We do not recover the counterpart in any images, and report upper limits in Table~\ref{tab:phot}.
In addition to observations with our programs, we collect optical and near-IR data from the GCNs and summarize them in Table~\ref{tab:phot}.

\subsection{Radio Observations}

We observed GRB\,250702BDE with the MeerKAT radio telescope \citep{Camilo2018, Jonas2018} under program SCI-20241101-FC-01 (PI Carotenuto) beginning on 2025-07-04 17:37:32 UT ($\delta t = 2.208$~days) at S-band (S4; central frequency 3.06\,GHz and 875\,MHz bandwidth), followed by a second observation on July 8th 2025 at 18:43:45 UT ($\delta t = 6.236$~days) at L-band (central frequency 1.28\,GHz and 856\,MHz bandwidth). We utilize PKS~J1939-6342 as bandpass and flux density calibrator and J1822-096 and J1908-201 as complex gain calibrators at 3.06 and 1.28 GHz, respectively, obtaining a total on-source time of 42 minutes in each band. We reduce the data with the \texttt{OxKAT} pipeline \citep{oxkat}, which performs standard flagging, calibration and imaging using \texttt{tricolour} \citep{Hugo_2022}, \texttt{CASA} \citep{CASA_team_2022} and \texttt{WSCLEAN} \citep{Offringa_wsclean}, respectively. For imaging, we adopt a Briggs weighting scheme with a $-0.3$ robust parameter for both frequencies, which yielded a beam of $10.9\arcsec \times 2.7\arcsec$ and rms noise of $9\,\mu$Jy\,beam$^{-1}$ at 3.06 GHz, and $8.1\arcsec \times 6.4\arcsec$ and $13\,\mu$Jy\,beam$^{-1}$ at 1.28 GHz. We clearly detect the radio counterpart of GRB\,250702BDE at the position of the optical counterpart at both frequencies (first reported in \citealt{gcn40985}). We fit for a point source in the image plane and report our results in Table~\ref{tab:radio}.
We additionally compile publicly reported millimeter and radio observations from MeerKAT, the Atacama Large Millimeter Array (ALMA; \citealt{gcn41059}), the James Clark Maxwell Telescope (JCMT; \citealt{gcn41061}, the Allen Telescope Array \citep{gcn41053} and the Karl G. Jansky Very Large Array (VLA; \citealt{gcn41053}) in Table~\ref{tab:radio}.

\begin{deluxetable*}{ccccccc}
\tabletypesize{\small}
\centering
\tablecolumns{7}
\tabcolsep0.1in
\tablecaption{Log of optical/near-IR observations. The time is given relative to the initial GBM trigger on GRB\,250702D.
\label{tab:phot}}
\tablehead {
\colhead {Date}		&
\colhead {$\delta$T (days)} &
\colhead {Telescope/Instrument} &
\colhead {Filter} &
\colhead {Magnitude} &
\colhead {Magnitude (host sub)} &
\colhead {Reference}}
\startdata 
2025-07-03 07:03:03 & 0.746  & VLT/HAWK-I & $H$ & $20.78 \pm 0.05$ & --- & This work \\
2025-07-03 07:39:40 & 0.771  & VLT/HAWK-I & $K$ & $19.36 \pm 0.02$ & 19.41 $\pm$ 0.02 & This work\\
2025-07-03 17:26:10 & 1.179  & WFST & $r$ & $>22.2$ & --- & \cite{gcn40943} \\
2025-07-03 21:52:47 & 1.364  & FTW/3KK & $r$ & $>23.6$ & --- & \cite{gcn40949} \\
2025-07-03 21:52:47 & 1.364  & FTW/3KK & $i$ & $>22.7$ & --- & \cite{gcn40949} \\
2025-07-03 21:52:47 & 1.364  & FTW/3KK & $J$ & $>21.7$ & --- & \cite{gcn40949} \\
2025-07-03 22:50:00 & 1.403  & NOT/StanCAM & $i$ & $>24.9$ & --- & This work \\
2025-07-03 23:45:36 & 1.442  & CAHA2.2/CAFOS & $z$ & $>21.4$ & --- & \cite{gcn40929} \\
2025-07-04 03:20:59 & 1.591  & VLT/HAWK-I & $K$ & $20.85 \pm 0.03$ & 21.05 $\pm$ 0.04 & This work\\
2025-07-04 03:36:09 & 1.602  & VLT/HAWK-I & $H$ & $22.39 \pm 0.06$ & --- &This work\\
2025-07-05 01:44:52 & 2.525   & GTC/HiPERCAM & $g$ & $>23.6$ & --- & This work\\
2025-07-05 01:44:52 & 2.525   & GTC/HiPERCAM & $r$ & $>23.0 $ & --- & This work\\
2025-07-05 01:44:52 & 2.525   & GTC/HiPERCAM & $i$ & $>22.6$ & --- & This work\\
2025-07-05 01:44:52 & 2.525   & GTC/HiPERCAM & $z$ & $>22.3$ & --- & This work\\
2025-07-05 03:07:37 & 2.582  & VLT/HAWK-I & $K$ & $21.49 \pm 0.05$ & 21.88 $\pm 0.08$ & This work\\
2025-07-05 10:14:00 & 2.878  & Keck/MOSFIRE & $K$ & $21.7\pm0.1$ &  --- &\cite{gcn41044} \\
2025-07-05 10:14:00 & 2.878  & Keck/MOSFIRE & $H$ & $>22.4$ &  --- &\cite{gcn41044} \\
2025-07-05 10:14:00 & 2.878  & Keck/MOSFIRE & $J$ & $>22.8$ &  --- &\cite{gcn41044} \\
2025-07-06 02:16:22 & 3.547   & GTC/EMIR & $Ks$ & $>21.5$ & --- & This work\\
2025-07-08 02:43:27 & 5.567   & GTC/EMIR & $Ks$ & $>22.0$ & --- & This work\\
2025-07-08 06:50:09 & 5.737  & VLT/HAWK-I & $K$ & $22.53 \pm 0.06$ & 24.17 $\pm$ 0.40 & This work\\
2025-07-15 01:40:27 & 12.52  & VLT/HAWK-I& $K$ & $22.82 \pm 0.08$ & --- & This work\\
2025-07-15 02:39:23 & 12.52  & \textit{HST}  & F160W & $24.80 \pm 0.30$ & --- & This work \\
\enddata
\tablecomments{Magnitudes were calibrated against UKIDSS nearby reference stars in Vega and then converted to AB by applying offsets of 1.900 ($K$-band) and 1.379 ($H$-band) \citep{hewett06}. Magnitudes are not corrected for Galactic extinction of $A_H=0.158$, $A_K=0.093$ \citep{SF11}.}
\end{deluxetable*}

\begin{deluxetable*}{cccccc}
\tabletypesize{\small}
\centering
\tablecolumns{6}
\tabcolsep0.1in
\tablecaption{Log of radio/sub-mm observations presented in this work or publicly available.  The time is given relative to the initial GBM trigger on GRB\,250702D.
\label{tab:radio}}
\tablehead {
\colhead {Date}		&
\colhead {$\delta$T (days)} &
\colhead {Telescope} &
\colhead {Frequency (GHz)} &
\colhead {Flux density (mJy/beam)} &
\colhead {Reference}}
\startdata 
2025-07-04 07:37:01 & 1.769 & Allen TA & 5     & $<1.1$ & \cite{gcn40979} \\
2025-07-04 18:08:44 & 2.208 & MeerKAT  & 3.06  & $0.12\pm0.01$ & This work \\
2025-07-06 18:53:41 & 4.239 & MeerKAT  & 1.28  & 0.1 & \cite{gcn41054} \\
2025-07-08 18:43:45  & 6.236 & MeerKAT  & 1.28  & $0.08\pm0.01$ & This work \\
2025-07-08 05:59:11 & 5.701 & VLA      & 10    & $0.49\pm0.05$ & \cite{gcn41053} \\
2025-07-08 21:05:00 & 6.331 & JCMT     & 350   & $<6.3$ & \cite{gcn41061} \\
2025-07-09 02:09:03 & 6.542 & ALMA     & 97.5  & 2 & \cite{gcn41059} \\
\enddata
\end{deluxetable*}

\section{Host galaxy}
\label{text:host}
The combination of the offset and measured $K$-band magnitude of the underlying galaxy enables us to calculate the probability of chance alignment ($P_{\rm cc}$). Following the methodology of \cite{bloom02}, but with number counts updated and appropriate for our $K$-band observations \citep{windhorst23}, we find $P_{\rm cc}\lesssim0.1\%$. We therefore conclude that, despite a location close to the Galactic plane ($b=-5.2$ degrees), the underlying galaxy is the host and that GRB\,250702BDE is an extragalactic event. 

The galaxy is clearly resolved in \textit{HST} images and visible as an extension in the HAWK-I images with good seeing. As the source is relatively extended and shows low surface brightness regions that overlap with foreground objects, accurate total photometry is particularly complex. Using an elliptical aperture of semi-major/minor axis 1.75\arcsec/0.65\arcsec, we measure magnitudes of $K(AB) = 20.78 \pm 0.05$ and F160W(AB) = $21.55 \pm 0.09$, and a half light radius of approximately 0.8\arcsec. Alternatively, using GALFIT \citep{peng10} and two Sersic functions, the best-fit model returns a mean half light radius of $0.9\pm 0.1$\arcsec, fully consistent with the elliptical aperture measurement. 

In the {\em HST} images the galaxy appears split into two separated regions. This may be due to some form of ongoing interaction, but is also reminiscent of strong dust lanes observed frequently from edge-on disc-like galaxies. The lane appears to extend across the nucleus of the galaxy in the direction of the near-IR counterpart; thus, if the transient lies within (or behind) the dust lane, dust extinction would provide a natural explanation for the observed extreme redness (Section~\ref{text:data:opt}).

No spectroscopic redshift for the galaxy is available at the time of writing. Existing detections, which are limited to F160W/$H$- and $K$-bands, are insufficient to derive a photometric redshift. However, the galaxy appears relatively large and well resolved, revealing substructure at the resolution of the {\em HST} observations, and enabling some constraints on its redshift. At $z=1$ for example, the observed $K$-band host magnitude would correspond to $M_z \sim -22.8$, substantially brighter than L$_{\star}$. At this redshift, the galaxy would also be much larger than GRB host galaxies \citep[e.g.][]{lyman17,schneider22}. Indeed, at such high redshifts it would be difficult to resolve dust lanes. We therefore believe it is more likely that the host of GRB\,250702BDE lies at $z<1$ (and most likely at $z<0.5$). We discuss additional constraints on the redshift in Section~\ref{text:agmodel}.

\section{A repeating, extragalactic gamma-ray burst}
\label{text:repeater}

Numerous properties of GRB\,250702BDE are unique or extreme when compared to those typically seen in GRBs. The first reported high-energy detections  pre-date the final GBM trigger by $>24$ hours (Section~\ref{text:xray}). Whereas most GRBs (as well as ultra-long GRBs and relativistic jetted TDEs) are brightest in the $\gamma$-rays within seconds to minutes of the trigger time, there were no $\gamma$-ray triggers from GRB\,250702BDE prior to $T_0$, despite the earlier detections by the {\em Einstein Probe} \citep{gcn40906}. Further, the most energetic outburst was the final one (GRB\,250702E) \citep{gcn40931}, with the fluence of the three bursts reported as 1.04 $\times 10^{-5}$, 1.24$\times 10^{-5}$ and 2.74 $\times 10^{-5}$ erg cm$^{-2}$. Hence, both in timescale and light curve morphology, GRB\,250702BDE seems to be substantially different from the bulk of the GRB population. 

The timing of the three triggers further exhibits interesting patterns. Whereas the time difference between the first and second GRBs is 2825\,s, the third event (GRB\,250702E) begins at near-integer time-step from this, corresponding to $4.0003$ times the 2825\,s interval (i.e., $(4\times 2825)+4$\,s later). This could be interpreted as a period of repetition in the GRBs with potential implications for the progenitor. If so, these three GRB correspond to the first, second and fifth ``beat'' of the period. No $\gamma$-ray triggers have been reported coincident with the third and fourth beat. Although plausible sub-threshold structure is visible within the $\gamma$-ray light curves (Figure~\ref{fig:fermi_prompt}), the complex background precludes any robust statements about the presence of possible emission at these times. We consider the possible physical implications of this potential periodicity in Section~\ref{sec:tde}.

Although there is no precise redshift available for GRB\,250702BDE, the presence of a relatively bright and extended host galaxy suggests the redshift is low ($z<1$; Section~\ref{text:host}), and our afterglow modeling (Section~\ref{text:agmodel}) favors a $z \sim 0.2$ origin. Using these constraints, we next consider how the energetics of the burst would appear at redshifts in the interval $0.1 < z < 1$. Unsurprisingly $E_{\gamma,\rm iso}$ varies dramatically across this broad redshift range. Using the integrated Konus-Wind flux for GRB\,250702BDE, the $E_{\rm iso}$ ranges from $1.4 \times 10^{52}$ erg ($z=0.1$) to $1.6 \times 10^{54}$ ($z=1.0$), which are both consistent with the bulk GRB population. Strikingly however, the joint spectrum of the three bursts as seen by Konus-Wind in the 20--1250\,keV range (observer frame) is well described by a power law \citep{gcn40914}, and indicates that the peak of the prompt $\nu F_{\nu}$ spectrum ($E_p$) lies at $\gtrsim1\,$MeV. Indeed, even attempts to fit typical GRB-like Band functions \citep{band93} to the spectra yield high $E_p$ of several hundred keV \citep{gcn40931}. Notably across the redshift range considered here, this lies substantially in excess of the $E_p$--$E_{\rm iso}$ relations \citep[e.g.][]{amati06}. The same is true, and indeed becomes more acute, if considering each individual burst (``D", ``B", ``E") separately. 

\section{Afterglow properties}
\label{text:agmodel}

\begin{figure}
\centering
\includegraphics[angle=0,width=\columnwidth]{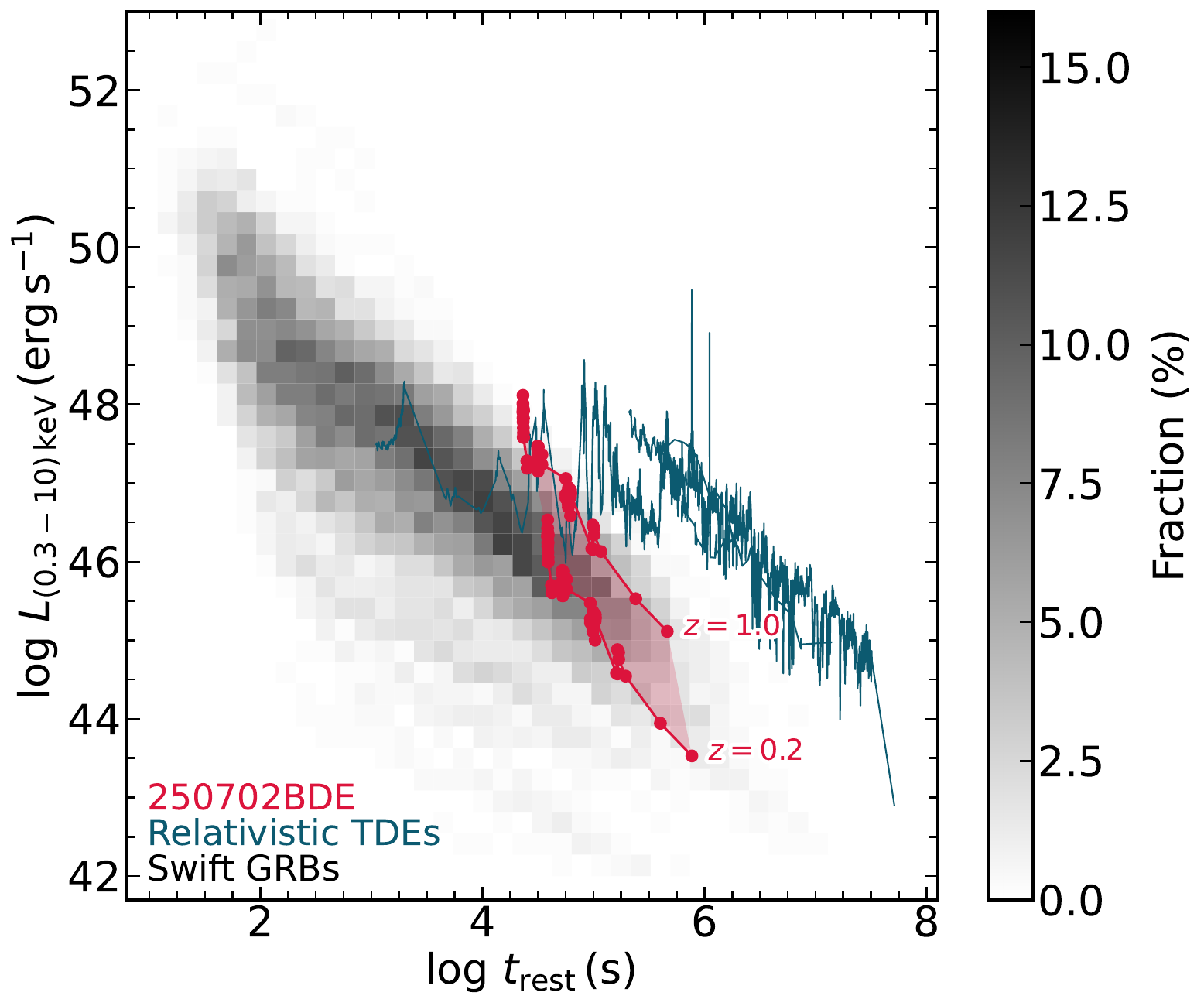}
\caption{Comparison of the X-ray light curve of GRB\,250702BDE with those of {\em Swift} GRBs (greyscale background) and relativistic TDE candidates, downloaded from the Swift Burst Analyser \citep{evans07, evans09, Evans2010a} and the Living Swift XRT Point Source Catalogue \citep{evans23} and processed as described in \citet{Schulze2014a}. At any reasonable redshift the luminosity of the event is comparable to those of GRBs, and GRB\,250702BDE is clearly faster and less luminous than the relativistic TDE candidates. This would be consistent with a lower-mass black hole involved in a disruption, or with a more normal GRB interpretation.}
\label{fig:schulze}
\end{figure}

\begin{figure*}
\centering
\begin{tabular}{cc}
\includegraphics[width=\columnwidth]{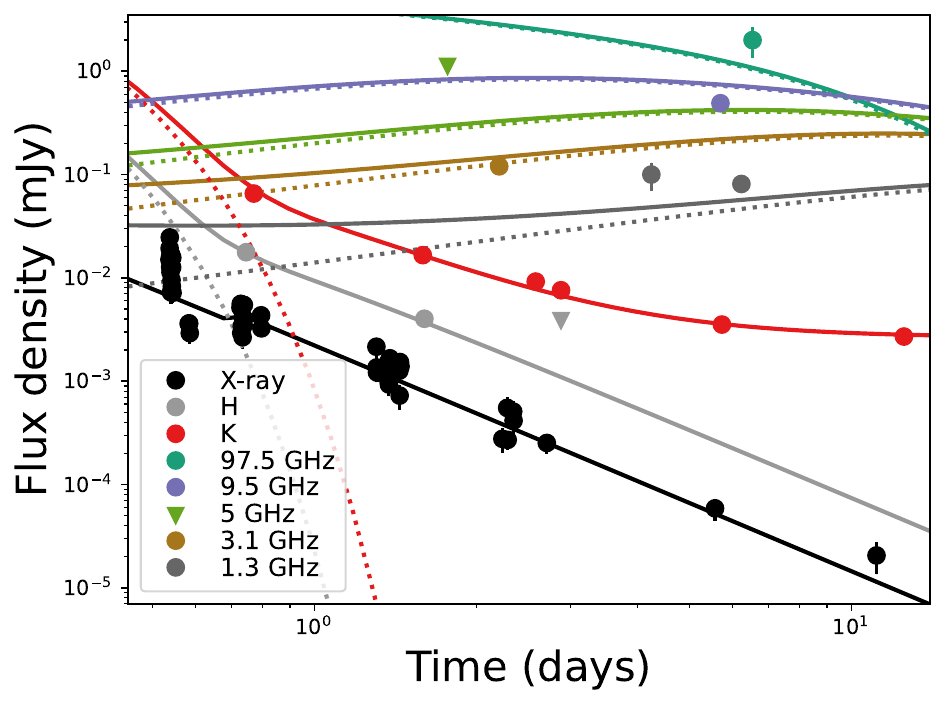}&
\includegraphics[width=\columnwidth]{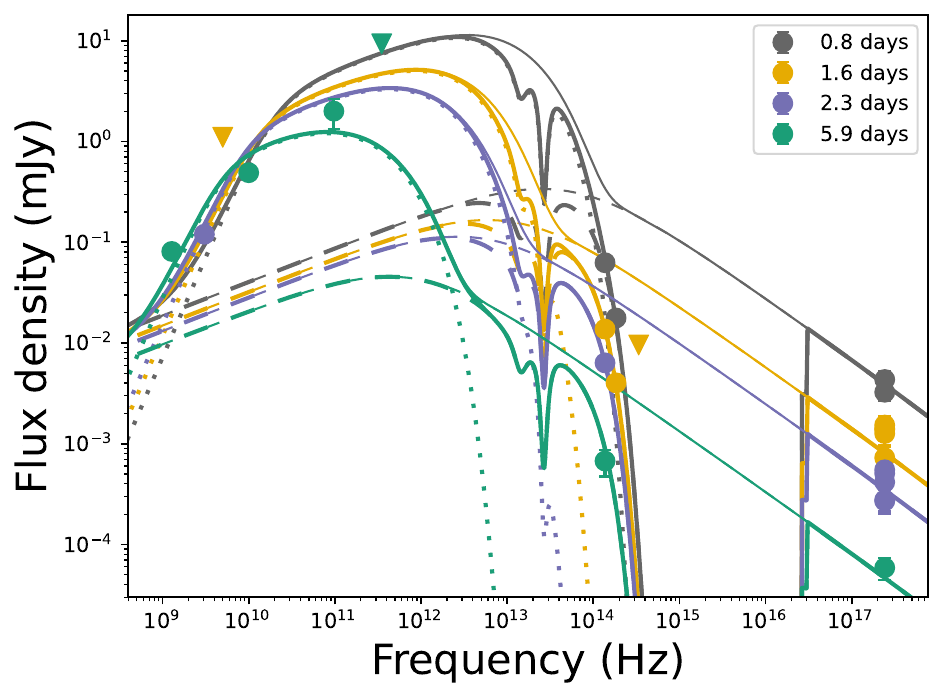}
\end{tabular}
\caption{Light curves (left) and spectral energy distributions (right) for the afterglow of GRB\,250702BDE spanning from $\approx0.5$ to 20 days at radio to X-ray wavelengths. The observations can be well explained by a standard synchrotron model comprising radiation (solid lines) from a forward (dashed) and reverse (dotted) shocks at $z\approx0.16$ and high host attenuation, $A_V\approx12$~mag (see Section~\ref{text:agmodel} for details and Appendix~\ref{appendix:afterglow} for model parameters).}
\label{fig:afterglow}
\end{figure*}

While the prompt emission seems unusual, the afterglow is more readily comparable to those of the typical GRB population. Across the range $0.1 < z < 1$, the X-ray and radio luminosity is consistent with GRB afterglows (e.g., Figure~\ref{fig:schulze}; \citealt{cf12,duplm+12}). Though the near-IR emission is unusually red, its absolute magnitude is also in keeping with those seen for other GRB afterglows \citep[e.g.,][]{kann24}. 

We now consider whether these observations are consistent within the framework of the standard GRB fireball model, where synchrotron afterglow emission arises from shocks produced in the interaction of a collimated, relativistic jet with the ambient medium (see Figure~\ref{fig:afterglow} left panel). The details of our modeling are discussed in Appendix~\ref{appendix:afterglow}. We keep the redshift as a free parameter in the range $0.1<z<12$ and find that forward shock (FS) emission fits the optical and X-ray observations well, provided $z\lesssim0.3$ and $A_V\gtrsim10$~mag. Higher redshift models, while requiring lower $A_V$ (due to the near-IR observations probing bluer rest-frame filters), push\footnote{This degeneracy maintains a high cooling frequency, $\nu_{\rm c}\gtrsim\nu_{\rm X}$, which is required for matching the de-reddened near-IR colors and the near-IR to X-ray spectral index at higher $z$ within our allowed constraint of $2<p<3$.} the outflow's isotropic-equivalent energy ($E_{\rm K,iso}$) beyond our maximum allowed value of $5\times10^{54}$~erg and are disfavored (see Figure~\ref{fig:agcorner} in the Appendix). The observed steep X-ray and optical light curves are further consistent with an early jet break ($t_{\rm jet}\lesssim0.5$~days). The radio observations in this framework require an additional component, e.g., a reverse shock (RS; Figure~\ref{fig:afterglow} right panel). In our modeling, we keep this component fixed (due to the paucity of radio data) and vary the FS parameters. Besides an extremely narrow opening angle ($\theta_{\rm jet}\approx0.4^\circ$, driven by the early jet break, high $E_{\rm K,iso}$, and low density), the inferred parameters are consistent with those previously inferred for GRBs (e.g., \citealt{lbt+14}). For our best-fit parameters, we infer a prompt efficiency of $\eta_\gamma\approx10\%$ and an FS Lorentz factor of $\Gamma_{\rm FS}\approx40$ at the time of the first detection, $\delta t\approx0.5$~days, also consistent with GRB afterglows \citep{lbt+14,gng+18}.

The high local host extinction is also consistent with the observed large X-ray absorption column, $N_{\rm H_X}\approx(1.1\pm0.2)\times10^{22}\,{\rm cm}^{-2}$ (corresponding to $N_{\rm H_X}\approx(8.6\pm2.0)\times10^{21}\,{\rm cm}^{-2}$ after subtracting the Galactic contribution;  \citealt{evans07,evans09}). We can formalize this by computing a redshift using the correlation between X-ray and dust column \citep[e.g.,][]{watson11}, $z=\left(\frac{2\times10^{21}\,\mathrm{cm}^{-2}\,\mathrm{mag}^{-1}}{N_{\rm H_X}(z=0)/A_V(z)}\right)^\frac{1}{2.4} - 1$ (Appendix~\ref{appendix:NH-z}). 
Using our inferred host galaxy $A_V(z=0)$ posterior and additional statistical uncertainties (Appendix~\ref{appendix:NH-z}), this gives a redshift of $z = 0.50\pm0.29$, consistent with the host properties (Section~\ref{text:host}) and the afterglow modeling (Appendix~\ref{appendix:afterglow}). One caveat here is that for classical GRBs, this ratio is an order of magnitude larger \citep[e.g.][]{schady10}, probably due to a large column density of ionized, dust-free gas close to the GRB site \citep{watson13}. Thus, if this was a classical GRB, then the redshift inferred above is a lower limit. 
This constraint provides further evidence the burst is not Galactic, because the metals providing the dust column would be detected in stronger X-ray absorption than is observed.

\section{Progenitor scenarios}

The key question relating to GRB\,250702BDE is the identity of its progenitor. As a series of long GRBs, does it arise from a stellar-scale core-collapse event as is apparently the case for the ultra-long GRBs \citep{greiner15}? Alternatively, given its unprecedented nature, does it arise from a potentially previously unseen kind of object? 
Here we consider several possible progenitor scenarios for GRB\,250702BDE, and contrast these against available observations. 

\subsection{A relativistic TDE}
\label{sec:tde}

A known source class that can produce high-energy emission on $\sim$hours to days-timescales is relativistic TDEs powered by super-Eddington accretion onto a massive black hole.
Highly variable $\gamma$-ray light curves similar to GRB\,250702BDE have previously been observed in relativistic TDEs, specifically Swift J1644+57 \citep[e.g.][]{burrows11,Mangano2016a,levan16} which triggered \textit{Swift}-BAT (via an image trigger) four times over two days with a comparable inter-trigger interval to GRB\,250702BDE. 
The high $\gamma$-ray variability, hard-to-soft X-ray spectral evolution, and variable (but non-flaring) hard X-ray properties are all consistent with the early behaviour of past relativistic TDEs \citep{Burrows2011a,Cenko2012a,Pasham2015a,Mangano2016a}. This similarity extends further with consistencies between the near-IR power law decay and self-absorbed radio emission of GRB\,250702BDE these past events \citep{levan16,Hammerstein25}.
Underneath the short term variability, the early X-ray light curve of relativistic TDEs typically follows a power law decay, sometimes consistent with the canonical $t^{-5/3}$ decay in the fallback rate, $\dot{m}_{\rm fb}$ \citep{rees1988}. However, this can be steeper and consistent with the decay ($\alpha_{\rm X}=-2.0\pm0.1$) we see in GRB 250702BDE due to factors we discuss below.

We consider three possible TDE scenarios for GRB\,250702BDE that produce sufficiently super-Eddington accretion rates to drive a relativistic jet: (1) a main sequence (MS) star disrupted by a supermassive black hole (SMBH, $\sim10^5-10^7 M_\odot$), (2) a MS disrupted by an IMBH, and (3) a WD disrupted by an IMBH. Notably, each of these is observationally rare, with only four MS-SMBH TDE candidates \citep[e.g.][]{Burrows2011a,levan11,bloom11,Cenko2012a,Brown2015a,andreoni23,Pasham2023a} and even fewer IMBH TDE candidates \citep[e.g.][]{jonker13,Jin25}, although the latter have also been considered for ultra-long GRBs \citep[e.g.][]{levan14,MacLeod2016}.

The non-nuclear location of GRB\,250702BDE in the host galaxy (Figure~\ref{fig:imaging}) makes the first scenario unlikely but still feasible if, e.g., the host is undergoing a merger. As this is difficult to assess at present, we disfavor but do not rule out this scenario.

Distinguishing between the second (MS-IMBH) and third (WD-IMBH) scenarios is harder, but may be possible if the periodicity between the triggers (Section~\ref{text:repeater}) is associated with the orbital period of the disrupted star (i.e., the $\gamma$-ray flares correspond to mass transfer caused by partial disruptions). In this case, we can compare the required circular orbital radius, the innermost stable circular orbit (ISCO) for a range of black hole masses and the Roche limits for a 1~M$_\odot$ MS and WD stars. This reveals that only WD-IMBH disruptions are consistent with the periodicity. The short term ($\sim 100$s) variability observed in the light curve is also easier to achieve if associated with precession, as suggested for Swift J1644+57 \citep{Krolik11}. In the TDE scenario, we thus favor the WD-IMBH explanation; however, we note that the multiple bursts may also be driven by a jet or accretion timescale unrelated to an orbital period (e.g., precession).

For a fiducial $10^4$~M$_\odot$ IMBH and 0.5 $M_\odot$ WD, assuming the periastron radius is equal to the tidal radius, an eccentricity of order $e\sim0.955$ is required. This could plausibly be achieved through Hills capture \citep{Hills88} and would require an impact parameter of $\beta \sim 2$ for the original binary \citep{Cufari22}.
The eccentricity inferred above can result in a steeper X-ray decline than the typical $t^{-5/3}$ decay rate \citep{Cufari22b}. We test this using the `frozen-in' approximation\footnote{We note that \citet{Cufari22b} do find significant differences between this analytical approximation and numerical simulations but these dominate at more extreme eccentricities.} with the $10^4 M_\odot$ IMBH, $0.5M_\odot$ WD and $e=0.955$ above. We fix the peak fallback time to the \textit{Fermi}-GBM trigger time of GRB\,250702E \citep{gcn40891} and, assuming an accretion efficiency of order 0.01 and $z\sim0.14$ to match the observed luminosity of Swift J1644+57, find a reasonable match to the X-ray light curve\footnote{While the jet physics involved are complex, if the jet luminosity is dominated by the accretion rate, the difference in slope could also be attributable to a full disruption ($\dot{m}_{\rm fb}\propto t^{-5/3}$) in Swift J1644+57 and a partial disruption ($\dot{m}_{\rm fb}\propto t^{-9/4}$) in other events including GRB\,250702BDE \citep{Nixon21,Eftekhari24}.}.

Overall, if GRB\,250702BDE is indeed a relativistic TDE, the data discussed here likely sit within the jet-dominated phase which is seen to last of order months to years in other events \citep{Zauderer13,Brown2015a,Pasham2015a,Eftekhari17,Eftekhari24}.

A distinguishing signature of a relativistic TDE is the X-ray emission entering a steep decline 
as the accretion rate drops below the Eddington limit and the jet turns off \citep{levan16,Pasham2015a,Eftekhari24}. Based on the analytical WD-IMBH model described above, we estimate this will occur at $\sim40$ days in the observer frame. We note that due to the degeneracy of the efficiency and redshift, the steep decline may occur $\sim$months to years later, particularly the IMBH is less massive than assumed. The radio counterparts of relativistic TDEs continue to rise for hundreds to thousands of days and may also show evidence for multiple components \citep{Zauderer13,Pasham2015a,Eftekhari17,Teboul23}. In this case, multiple components could be expected in a similar fashion to Swift J1644+57 \citep[e.g.][]{Teboul23} while the rise could be curtailed at relatively early times if the jet shuts off in line with our prediction. Long-term X-ray and radio monitoring is therefore crucial to evaluating this explanation.

\subsection{An atypical collapsar GRB}
The majority of long-duration GRBs appear to arise from the core collapse of massive stars \citep[e.g.][]{hjorth03,stanek03}, although a substantial minority now also seem to arise from compact object mergers \citep[e.g][]{rastinejad22,troja22,yang22,levan24,yang24}. In principle, the duration of the GRB reflects the lifetime of the central engine {\em after} the jet has pierced the star \citep{bromberg13}. The time for the jet to traverse the star is a few seconds for a compact object merger, up to a few hours for the collapse of giant stars \citep{levan14}. In the case of GRB 250702BDE the combination of duration (too long), morphology (starting faint and getting brighter), and prompt energy correlations (away from the $E_p - E_{iso}$ relations) challenge the interpretation of GRB 250702BDE as a core 
collapse event.

Although several lines of evidence point away from a collapsar, it should also be noted that
several others are entirely in keeping with a collapsar origin. These include the overall energetics, afterglow properties, and the location within the host galaxy. Although the afterglow is highly extinguished, there is strong precedent for this in the dark GRB population \citep{svensson12,melandri12,perley13,schroeder22}.

As a rare event it is also possible that unusual and rare circumstances are at play in GRB 250702BDE that do not impact the general GRB population, including GRBs originating at unusual phases of stellar evolution. This might include explosions within a common envelope \citep[e.g.][]{schroder20} that could create complex circumstellar media into which the jet propagates and the $\gamma$-rays are scattered, the collapse of particularly massive stars formed dynamically in extremely dense stellar clusters \citep[e.g.][]{pz04}, or even stellar collapse inside a very tight binary containing a second compact object, in which complex fallback processes impact the resulting light curves \citep[e.g.][]{church12}. 
Hence, while clearly exhibiting a set of prompt properties inconsistent with those of GRB-supernovae seen before, these do not automatically rule out a stellar collapse origin. Indeed, such origins have even been suggested for the population of events now generally interpreted as relativistic TDEs \citep{quataert11}.

For a moderately low redshift it may ultimately be possible to observe any associated supernova, although the extinction will require $K$-band observations where supernovae are unfortunately not especially bright. At $z \sim 0.2$ we might expect a supernova near-IR peak around $K\sim22$~mag. Ground-based near-IR observations may be sensitive to supernovae even in cases where $A_K > 2$, and {\em JWST} could probe to very high extinctions ($A_K > 10$). For comparison, our afterglow modeling indicates $A_K\sim1.7$~mag, suggesting that a deep ground-based supernova search may yet be viable.

\subsection{A GRB with dust echoes}
The repeating nature of GRB\,250702BDE shares some superficial similarities with GRB dust echo models \citep[in which the prompt emission is Compton scattered by circumstellar material, e.g.][]{Dermer1991,madau2000,Moran2005,heng2007,shao2008}. The $\sim$1 hour-long timescale between peaks corresponds to an extremely compact physical scale of $\sim$10 AU for the scattering surfaces. In this interpretation, the distinct peaks require a specific configuration of the circumstellar medium, namely dense, thin shells \citep{madau2000}. Although the event apparently occurred in a dusty environment, the spatial scale implied by these timescales is small enough that dust destruction by the GRB radiation field should be significant \citep{heng2007}, decreasing the opacity and reducing the luminosity of the echoes. This is in contrast with GRB\,250702BDE which has similar (or even increasing) peak count rates in each of the three triggers (Figure \ref{fig:fermi_prompt}). On the other hand, the different distances associated with different parts of the scattering shells would naturally extend the prompt duration. Furthermore, the spectrum of the scattered prompt emission is expected to be significantly attenuated at energies above $\sim$100\,keV and to soften over time \citep{Dermer1991,sd07}. This is at odds with the high peak energy (and increasing spectral hardness) seen in the {\it Fermi}-GBM triggers. 

\subsection{A (lensed) high-z GRB}
The very red $H\,-\,K$ colour could be explained by the presence of the Lyman$-\alpha$ break in the $H$-band, placing the burst at $z \sim 12$. In this case, the foreground galaxy would either be a chance alignment, or more likely (see below) acting as a lens for the GRB. GRBs spawned from first generation (population III) stars are suggested to be both extremely energetic, $E_{\gamma,{\rm iso}}\gtrsim 10^{55}$\,erg, and long-lasting, $t_{\rm dur} \gtrsim 10^3\times(1+z)$\,s because of their extended envelopes \citep{Yudai2011}. At $z\sim12$, the isotropic energy of GRB would be $\sim 10^{56}$\,erg, an extreme value\footnote{Currently the highest isotropic energy measured for a GRB is $\sim\,2\times10^{55}$\,erg for GRB\,221009A \citep[e.g.][]{burns23, malesani23,atteia25}.}, but rendering the energetics consistent with the $E_p$ - $E_{\rm iso}$ relation (Section~\ref{text:repeater}). Alternatively, the long duration of the event could be due to lensing, with the same burst repeating multiple times. The concept of repeating GRBs as lensed events stretches back to some of the first suggestions of extragalactic origin \citep[e.g.,][]{pac86}, and a handful of claimed lensed GRBs have been suggested in the literature \citep[see][for a review]{levan2025}. 

However, while such a high-$z$ event would be of extreme importance, our X-shooter spectroscopy shows the source to be red across the spectral window, without a strong break. Hence, a high-$z$ event (lensed or not) would also need to be intrinsically red (and hence its afterglow even brighter allowing for this extinction). Furthermore, although the signal to noise of the prompt emission is not high, the morphology of the burst light curves appears to be different, disfavoring the lensing origin.

\section{Conclusion}

We have presented new, multi-wavelength observations of a superlative series of associated GRBs, GRB\,250702BDE. Our observations reveal a rapidly-fading, multi-wavelength counterpart embedded in a galaxy with a complex and asymmetric morphology. We identify this galaxy as its host, and conclude that GRB\,250702BDE is an extragalactic event. 

GRB\,250702BDE is observationally unprecedented in both its timescale, morphology, and the onset of X-ray photons prior to the initial GRB trigger \citep{gcn40906}. 
In addition, we find a striking, near-integer time-step between the GRB outbursts, suggesting (although not proving) possible periodicity in the events.

We find that a standard afterglow forward and reverse shock model can explain the multi-wavelength light curves of the burst counterpart. The model favors high local host attenuation ($A_V \approx 11$~mag) and a low redshift origin, leaving open several possible progenitor theories. We explore several models for this unusual event, including a relativistic TDE, a rare, non-standard collapsar, a typical collapsar which produces GRB dust echoes, and a lensed, high-redshift event. We find that an atypical collapsar and a possibly repeating WD-IMBH TDE are most compatible with current available information.

Looking forward, high-resolution observations of GRB\,250702BDE's local environment, continued X-ray and radio monitoring of its counterpart, and a deep near-IR supernova search are promising routes to further constrain its origins. Future detections of such events, especially those exhibiting periodicity, offer the chance to constrain their rates and decipher their origins.

\section*{Acknowledgments}
Based on observations collected at the European Southern Observatory under ESO programme(s) 114.27PZ. This work is based on observations made with the NASA/ESA Hubble Space Telescope. The data were obtained from the Mikulski Archive for Space Telescopes at the Space Telescope Science Institute, which is operated by the Association of Universities for Research in Astronomy, Inc., under NASA contract NAS 5-03127 for JWST. These observations are associated with program 17988.
Based on observations made with the Gran Telescopio Canarias (GTC), installed at the Spanish Observatorio del Roque de los Muchachos of the Instituto de Astrofiscica de Canarias, on the Island of La Palma under programmes GTC1-24ITP (PI Jonker) and GTC MULTIPLEC3C-25A (PI Mata Sanchez)
Based on observations made with the Nordic Optical Telescope, owned in collaboration by the University of Turku and Aarhus University, and operated jointly by Aarhus University, the University of Turku and the University of Oslo, representing Denmark, Finland and Norway, the University of Iceland and Stockholm University at the Observatorio del Roque de los Muchachos, La Palma, Spain, of the Instituto de Astrofisica de Canarias. The NOT data were obtained under program ID P71-506 (PI: Daniele B. Malesani, Johan P. U. Fynbo, Dong Xu).
The MeerKAT telescope is operated by the South African Radio Astronomy Observatory, which is a facility of the National Research Foundation, an agency of the Department of Science and Innovation. This work has made use of the ``MPIfR S-band receiver system" designed, constructed and maintained by funding of the MPI f\"ur Radioastronomy and the Max Planck Society.
AMC and LC acknowledge support from the Irish Research Council Postgraduate Scholarship No. GOIPG/2022/1008. P.G.J. J.S.S., J.Q.V., and M.E.R.~are supported  by the European Union (ERC, Starstruck, 101095973, PI Jonker). Views and opinions expressed are however those of the author(s) only and do not necessarily reflect those of the European Union or the European Research Council Executive Agency. Neither the European Union nor the granting authority can be held responsible for them. BS and SDV acknowledge the support of the French Agence Nationale de la Recherche (ANR), under grant ANR-23-CE31-0011 (project PEGaSUS). AS acknowledges support by a postdoctoral fellowship from the CNES. J.C.R. acknowledges support from the Northwestern Presidential Fellowship.  This work was supported by a research grant (VIL54489) from VILLUM FONDEN. RS acknowledges Leverhulme Trust grant RP-2023-240. DBM, DW, and ASn are funded by the European Union (ERC, HEAVYMETAL, 101071865). The Cosmic Dawn Center (DAWN) is funded by the Danish National Research Foundation under grant DNRF140.

\appendix
\section{Afterglow Modelling}
\label{appendix:afterglow}
For our modeling, we download the {\em Swift}-XRT count-rate light curve and convert to flux density at 1\,keV using the spectral parameters reported on the {\em Swift} website\footnote{We use $\Gamma=1.6$ and a counts-to-flux conversion factor of $9\times10^{-11}{\rm erg\,cm^{-2}\,ct^{-1}}$. The parameters reported on the website may drift following continued monitoring.}. We use the NIR fluxes unsubtracted for host emission and incorporate a constant component at $K$-band to account for host contamination. Our modeling framework is described in detail elsewhere \citep{lbz+13,lbt+14}, but in summary, we use the standard analytical model of \cite{gs02} with allowance for dust extinction in the Milky Way and in the host galaxy (following a Small Megallanic Cloud dust law), radio scintillation, inverse-Compton cooling, and Klein-Nishina corrections \citep{pei92,gn06,ml24}. The free parameters of our FS model, along with their best-fit values, are the electron energy index ($p$), the fraction of post-shock energy in accelerated electrons ($\epsilon_{\rm e}$) and in the magnetic field ($\epsilon_{\rm B}$), $E_{\rm K,iso}$, the density parameter ($A_*$) for a wind-like environment, $t_{\rm jet}$, $F_{\nu,\rm host,K}$, and the redshift, $z$. We additionally incorporate an RS component with the following fixed locations of the three spectral break frequencies, $\nu_{\rm a}\approx2\times10^{12}$\,Hz (self-absorption), $\nu_{\rm m}\approx3\times10^{18}$\,Hz (peak), and $\nu_{\rm c}\approx8\times10^{18}$\,Hz (cooling), and peak flux, $F_{\nu,\rm m,RS}\approx5\times10^{4}$~mJy, all at a fiducial $\delta t=10^{-3}$~days, chosen to approximately match the radio spectral energy distributions without over-predicting the near-IR data. 

\begin{figure}
    \centering
    \includegraphics[width=\columnwidth]{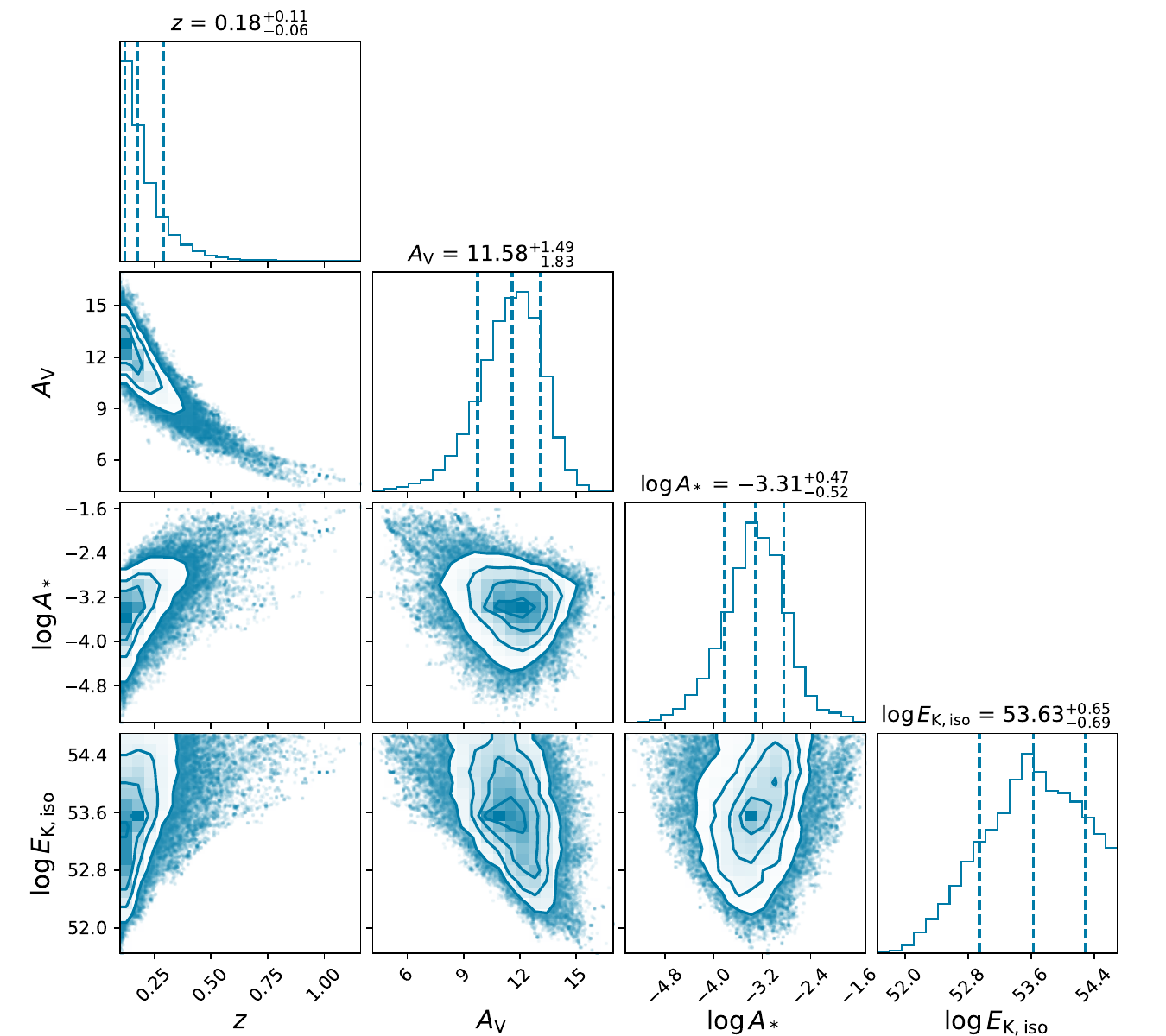}
    \caption{Correlation plots and marginalized posterior density functions for four key free parameters in the afterglow model (Section~\ref{text:agmodel} and Appendix~\ref{appendix:afterglow}), demonstrating that low-redshift $z\lesssim0.3$ and high-extinction ($A_V\gtrsim10$~mag) models are favored by the data under the modeling constraints employed here.}
    \label{fig:agcorner}
\end{figure}

We sample the parameter space using a Markov Chain Monte Carlo with \texttt{emcee} \citep{fmhlg13}. We run 512 walkers for 2\,000 steps and discard the first 30 steps as burn-in. We plot correlation contours for four parameters of particular interest in Figure~\ref{fig:agcorner}. 
The parameters of the highest likelihood model are $p\approx2.2$, $\epsilon_{\rm e}\approx0.95$, $\epsilon_{\rm B}\approx1\times10^{-2}$, $E_{\rm K,iso}\approx4\times10^{53}$~erg, $A_*\approx8\times10^{-4}$, $t_{\rm jet}\approx0.1$~days, $F_{\nu,\rm K,host}\approx2.9\times10^{-3}$\,mJy, and $z\approx0.16$. For this model (plotted in Figure~\ref{fig:afterglow}), the FS break frequencies and peak flux density at 1~day are $\nu_{\rm a}\approx7\times10^{5}$~Hz, $\nu_{\rm c}\approx3\times10^{20}$~Hz, $\nu_{\rm m}\approx3\times10^{13}$~Hz, and $F_{\nu,\rm m,FS}\approx0.5$~mJy. The outer break frequencies are unconstrained by the data in this model, resulting in degeneracies between the model parameters; however, these degeneracies are suppressed by the additional requirement of $\epsilon_{\rm e}+\epsilon_{\rm B}<1$, reducing the available parameter space and tightening the posterior in this case. 
While the data only constrain $t_{\rm jet}\lesssim0.5$~days (the first {\em Swift}-XRT detection), the smoothing used in the model light curves pushes the best-fit $t_{\rm jet}$ earlier. We infer a very small opening angle, $\theta_{\rm jet}=9.7^\circ \left[A_*/E_{\rm K,iso,52}(1+z)\right]^{1/4}\approx0.4^\circ$, which is lower than inferred for any afterglow previously \citep{lbm+15}, although perhaps not impossible to achieve (e.g., \citealt{sbat20}); nevertheless, we acknowledge this as a challenge that warrants further investigation beyond the scope of this work. Finally, we note that the RS parameters are simply chosen to roughly match the radio data; a more complete analysis of the RS (and the FS) will require the full multi-wavelength (and in particular, radio) dataset. We defer such analysis to future work.  

\section{X-ray column and redshift}
\label{appendix:NH-z}
Another constraint on the redshift can be obtained by comparing the X-ray absorbing column density to the optical/near-IR extinction. In the Galaxy, the X-ray absorbing column density, $N_{\rm H_X}$, is strongly correlated with the dust column \citep[e.g.][]{watson11}, and is expected to be strongly correlated for most galaxies, as the ISM X-ray absorption is due to metals, while dust is strongly correlated with the total metal column density. The ratio is about $N_{\rm H_X}/A_V\simeq2\times10^{21}$\,cm$^{-2}$\,mag$^{-1}$ \citep[e.g.][]{watson11,guver09,zhu17}. The extinction and X-ray absorption estimates scale with redshift. For extinction, this is $A_V(z) \simeq A_V(z=0)\times(1+z)^{-1.4}$, depending on the extinction curve. The X-ray absorption scales as $N_{\rm H_X}(z) \simeq N_{\rm H_X}(z=0)\times(1+z)^{2.4}$ \citep{campana14}. These work in opposing directions, such that $\frac{N_{\rm H_X}(z)/A_V(z)}{N_{\rm H_X}(z=0)/A_V(z=0)} = (1+z)^{3.8}$. If we assume the ratio $N_{\rm H_X}(z)/A_V(z) = 2\times10^{21}$\,cm$^{-2}$\,mag$^{-1}$, then we can infer the approximate redshift:  $z=\left(\frac{2\times10^{21}\,\mathrm{cm}^{-2}\,\mathrm{mag}^{-1}}{N_{\rm H_X}(z=0)/A_V(z=0)}\right)^\frac{1}{3.8} - 1$. If $A_V(z)$ is known (e.g., in our case from afterglow modeling), then the corresponding expression is $z=\left(\frac{2\times10^{21}\,\mathrm{cm}^{-2}\,\mathrm{mag}^{-1}}{N_{\rm H_X}(z=0)/A_V(z)}\right)^\frac{1}{2.4} - 1$. In practice, the X-ray absorption-to-extinction ratio and power law index will have uncertainties, for which we assume 10\% and $\pm0.1$, respectively.



\end{document}